\documentclass{PoS}

\title{``Light''  Higgs and warped  models: Case for a GIGANTIC INTERNATIONAL HADRON COLLIDER}

\ShortTitle{Higgs (125), warped models \& GIHC}

\author{\speaker{Amarjit Soni}\thanks{I want to thank Hooman Davoudiasl for many useful conversations.}\\
        Author affiliation\\High Energy Theory Group, Brookaven National Lab, Upton, NY 11973, USA 
        E-mail: \email{adlersoni@gmail.com}}

\abstract{The LHC seems to have  made a monumental discovery, Higgs-like particle of mass around 125 GeV 
          with properties akin to a Standard Model Higgs. In the context of a warped theory of flavor, which is theoretically very attractive,  this suggests Kaluza-Klein particle masses are likely to be above 10 TeV except possibly for a radion. The interpretation of the SM-like Higgs from the perspective of other interesting beyond the SM scenarios is also likely that the relevant scale is higher than accessible to the LHC. In light of these developments, deeper understanding of flavor and other fundamental issues requires a gigantic international hadron collider [GIHC] perhaps with cm energy of $\approx$ 100 TeV~\cite{2talks}. It is suggested that a {\it global effort} should be made for constructing this machine for resolving 
many questions that SM cannot answer.}

\FullConference{36th International Conference on High Energy Physics,\\
		July 4-11, 2012\\
		Melbourne, Australia}

\begin{document}

\section{Introduction}

Discovery of a Higgs-like object of mass around 125 GeV by the LHC experiments~\cite{Higgs}
in 2012 is quite a monumental one: its bound to prove as a watershed in our
understanding of the workings of Elementary Particle Physics.  So far the properties of this particle seem rather similar to those of a Standard Model [SM] Higgs~\cite{EWMoriond_2013}. Interpreted in the context of one  model of new physics that has compelling appeal, especially as a Theory of Flavor, namely warped models of flavor and hierarchy, this seems to imply that Kaluza-Klein resonances that are predicted in these models are
 over 10 TeV~\cite{ATZ10,GHN11,CCGHN12}.  This means except possibly  for a light (few hundred GeV) scalar, called radion~\cite{GW99}, in these models no other signal for new physics is likely to appear at the LHC~\cite{DMS12}. 

Actually, well known constraints from kaon mixing~\cite{MB07,BBS82} had already been suggesting rather strongly that KK-masses are quite difficult to be less than 10 TeV. So,  although electroweak precision constraints can be overcome by enforcing custodial symmetry and thereby the KK-scale lowered to around 3 TeV, the resulting set up needs some degree of tuning to satisfy kaon-mixing constraints unless the scale is above approximately 10 TeV.

Note that even with a KK-scale of 10 TeV, warped models of flavor score a big success in that the traditional flavor constraints of around $10^4$ TeV get lowered to around 10 TeV. Moreover, these models then no longer need imposition of custodial symmetry; after all this idea, clever as it is, has to introduce additional degrees of freedom and renders the models more intricate. Of course, this simplicity at 10 TeV (over 3 TeV)  comes at an expense as
the emergence of the electroweak scale out of 10 TeV is regarded as needing tuning of $O(10^{-3})$ range,
so this tuning  is worse compared to the case when KK masses are around 3 TeV by about one order of magnitude.
It is not clear, at least to this author, how serious an issue this is.  This extent of tuning is clearly a far cry from the original
problem of  tuning of O($10^{-34}$) if the relevant scale is O($10^{19}$) GeV and  Randall-Sundrum (RS) ideas  were not invoked.

It is important to remind ourselves that even if the interpretation of the Higgs-like particle as being largely SM-like gets confirmed there are still
numerous unanswered questions. The search for answers will undoubtedly  require a new high energy collider.
In the late 80's and 90's considerable effort was mounted towards a pp machine at 40 TeV (the SSC).
The machine was abadoned only because of budgetary and political concerns not for any technical reasons.
It stands to reason then that over 25 years later, a  machine at (say) 100 TeV energy should be feasible
in so far as technical know- how is concerned. Such a machine will require a genuine international effort and should be based on
recognizing the geo- political realities of the day.

\section{  Warped models of hierarchy and flavor}

Assuming the recently discovered particle at 125 GeV is confirmed to be  (mostly) a SM-like Higgs particle,  it will still
require a mechanism to stabilize its mass against large radiative corrections. One very attractive way to do this
was suggested by Randall and Sundrum~\cite{RS1} invovlving the notion of a warped extra dimension. As is well knowm, with this idea this particular
weak-planck hierarchy is addressed  by putting the Higgs on the infa-red (IR) or TeV brane.  

Interesting as it already  is,  the RS idea of warped space is even more suitable as a geometric theory of flavor.
This is realized by putting the SM gauge fields as well as fermions in the 5D RS bulk
resulting in a remarkable framework that can address simultaneously both hierarchy and flavor puzzles~\cite{DHR_2000,
AP_2000,GN_2000}

Here are some of the highlights of this picture. The huge disparity in the oberved fermion masses is naturally
accomodated as in the underlying 5D theory the corresponding localization parameters are all of O(1)~\cite{MN_EWMoriond09}. The fermion masses 
are of course an input and the RS-picture can not claim to predict them but what is remarkable is that 
after the masses are put in by hand,
it leads to a natural understanding of the severe suppresion in FCNC amongst the light quarks.
Indeed despite tree-level FC interactions involving KK-gluons (resulting from rotation from interaction to mass basis),
these FC currents    are suppressed roughly at the same level as the FC-loops in the SM so that an RS-GIM mechanism arises~\cite{APS041,APS042}. As in the SM all the FCNCs are driven primarily by the heaviness of the top quark~\cite{APS06}. Once the KK
masses are taken to be $ \gtrsim $ 10 TEV  even the severe constraints imposed by the smallness of $\Delta m_K, \epsilon_K$, from neutral kaons,
can be accomodated  with little or no tuning~\cite{Blanke_et_al_08,Cassagrande_et_al_08}. 

While the RS idea is extremely attractive as it can provide  simultaneous resolution to Planck-EW hierrachy and also an understanding of the  flavor puzzle, it is based on a strong assumption that warping extends over many orders of magnitude from the EW scale to the Planck. In the absence of any compelling phenomenological constraint and/or experimental evidence  and since the RS-set up provides an interesting theory of flavor by itself, it may be worth while to consider if   RS is a viable theory of  flavor alone. In that  more modest scenario\cite{DPS08}, called ``Little Randall Sundrum (LRS)", a volume-truncated RS background is used only to address the hierachy between the EW (IR)  scale of O(0.1 TeV) to  the flavor (UV) scale around $10^3$ or $10^4$ TeV\cite{DPS08,MB_et_al_09}.  Even in this modest set up constraints from $K - \bar K$  require $m_{KK} \gtrsim$ 10 TeV.  

Thus the bottom line is that whether we have RS or LRS,  flavor constraints require $m_{KK} \gtrsim 10$ TeV.
But once KK particles are that heavy,   the ``bonus" is that EW constraints may be  automatically  satisfied and there may not be  a compelling
need for imposing custodial symmetry~\cite{custodial}. This means that the  theoretical setup is more economical.
Of course this comes with a price-tag that tuning of O($10^{-3}$) is necessary but it is unclear
 if this should be considered  that big a drawback.  Indeed, it should be recognized that this order of tuning
is drastically different from what is needed to address the original EW-Planck hierarchy, requiring
tuning to the level of $\approx 10^{-34}$!.  It is simply not obvious that nature would choose to remove
this remaining O($10^{-3}$) tuning at the cost of making the theory more complicated by extending SU(2)
to SU(2)XSU(2)XU(1) and by
 the introduction
of new degrees of freedom to enforce, in the example of RS models, custodial symmetry~\cite{custodial}.

In the past several years considerable amount of effort was put into studying signals at the LHC for the lightest  KK particles, such 
as the gluon~\cite{kkglu,kkglu2},the  graviton~\cite{kkgraviton1,kkgraviton2,kkgraviton3} and  gauge bosons that RS/LRS  models  predict ~\cite{Shri1}. These studies have  shown that KK-particles
with masses greater than about three TeV will be difficult to see at the LHC . Thus if KK masses are  constrained to be above $\approx$ 10 TeV as suggested above,  then the varification of RS models   will necessitate colliders with energy much greater
than at the CERN-LHC.

\begin{figure}
\begin{center}
\includegraphics[width=10cm,angle=90]{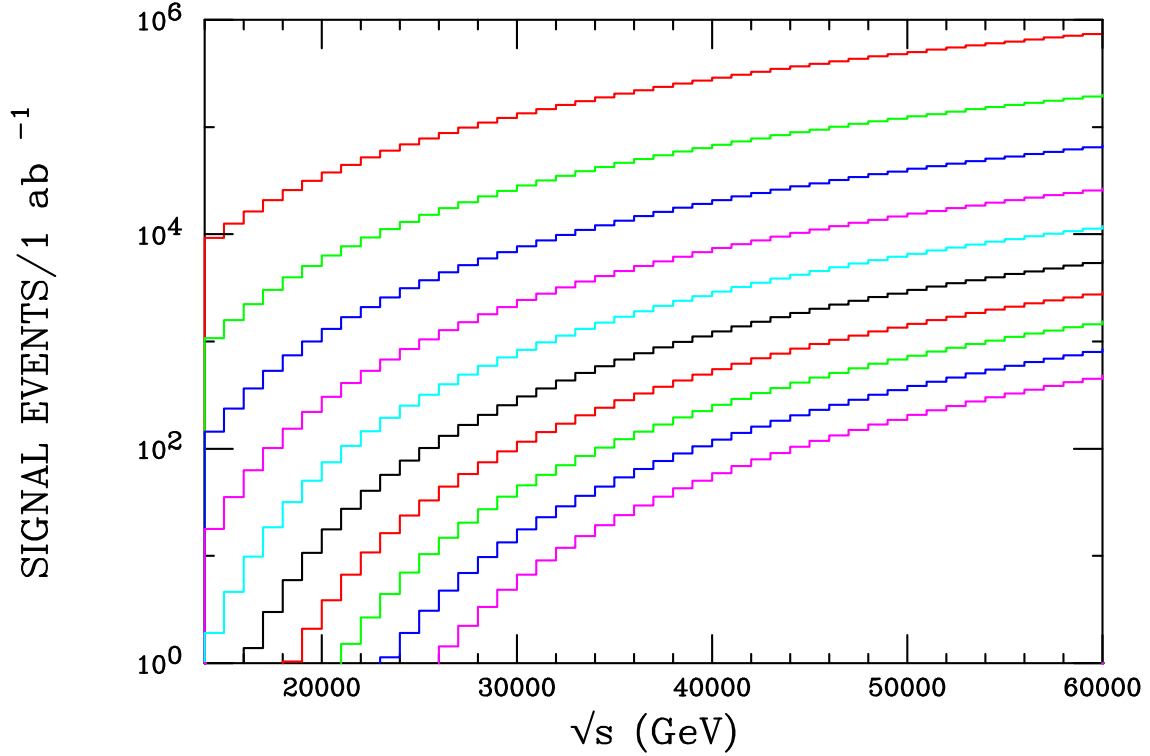}
\caption{ Expected rate for KK gluon $\to$ top pairs 
is shown versus collider energy for KK glu masses from
3 to 12 TeV. Branching fractions and efficiencies
are neglected. For cuts used and other details see the original ref.~\cite{DRS07}.
}
\label{fig:gluKKreach_S}
\end{center}
\end{figure}

 Fig.~\ref{fig:gluKKreach_S}) is  reproduced from ~\cite{DRS07}. This fig gives an indication of the expected signal
and its significance for KK gluon production and its detection via its decays to  $ t\bar t$.  Study includes some appropriate cuts to enhance signals. The maximum collider energy studied there is 60 TeV.  This fig  suggests that 
the next generation of supercollider with energy around 100 TeV could be quite effective in directly searching for warped resonances with masses around 10-20 TeV.

\begin{figure}
\begin{center}
\includegraphics[width=10cm]{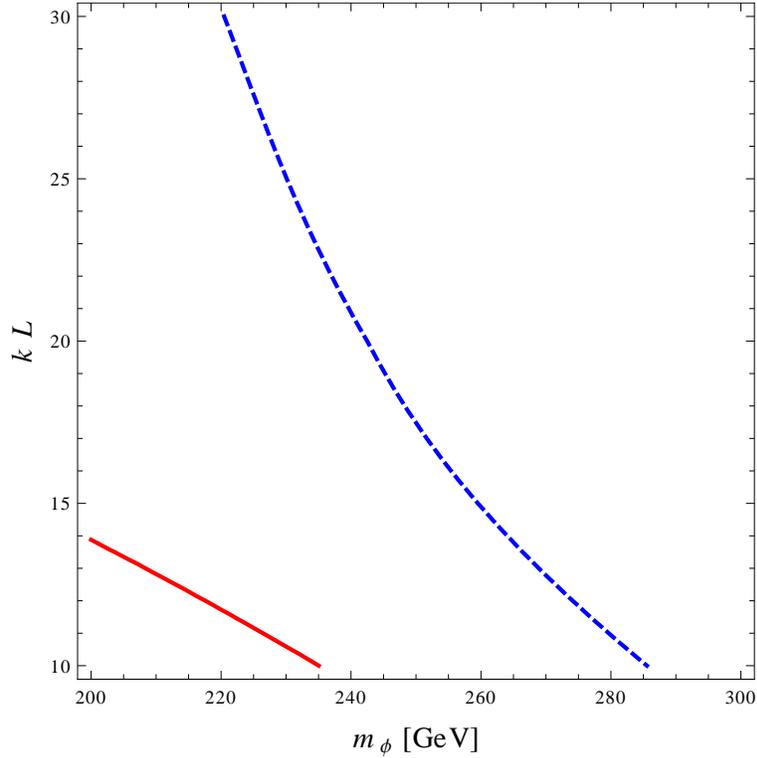}
\caption{ The $3\sigma$ (dashed)  and $5 \sigma$ (solid) contours, in the $(m_\phi,kL)$ plane, for $\phi\to W^+ W^- \to l^+ l^- \nu \bar \nu$ at the LHC with $100/fb$ at 14 TeV, with $\Lambda_{\phi}=10 $ TeV;  here $m_\phi$ is the radion mass, $10 \lesssim kL \lesssim 30$ covers the range  from LRS to RS and $ \Lambda_{\phi}$ is the characteristic KK-scale. Fig taken from~\cite{DMS12} where further details may be found.}
\label{fig:reach_m_kL}
\end{center}
\end{figure}

If this simpler picture ({\it i.e.} without including additional particles to impose custodial symmetry) of RS-models with 
$m_{KK} \gtrsim 10$ TeV is correct then the only signal of a warped underlying theory that we may see at the LHC is
from a radion as in these models it is expected to have a mass of several hundred GeV to $\approx$ a TeV.

\begin{figure}
\begin{center}
\includegraphics[width=10cm]{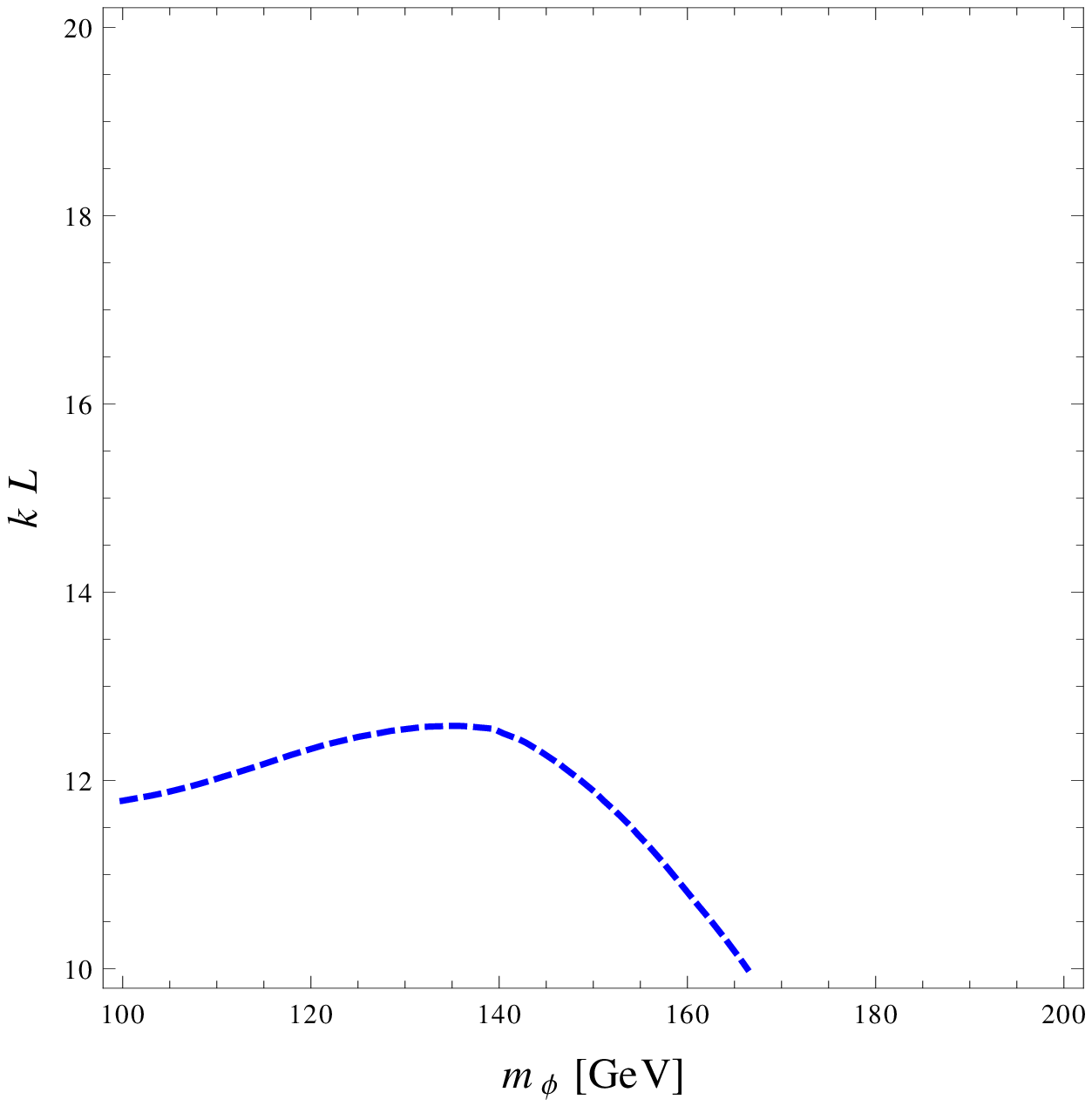}
\caption{ The $3\sigma$ contour, in the $(m_\phi,kL)$ plane, for $\phi\to\gamma\gamma$ at the LHC with $100/fb$  at 14 TeV, with $\Lambda_{\phi}=10$ TeV; from ~\cite{DMS12}.  }
\label{fig:diphoton}
\end{center}
\end{figure}

Feasibility of such signals is examined in ~\cite{DMS12} and  Fig.~\ref{fig:reach_m_kL} and
 Fig.~ \ref{fig:diphoton}   are from that study.
From these figs  the reach of the LHC at 14 TeV  for the detection of the radion decays via WW and $\gamma \gamma$ 
 are shown. It seems that if the radion is relatively light and $\lesssim$ about 250 GeV, the LHC may have a chance of 
seeing it, though appreciable fraction of the allowed parameter space, for larger masses,  is unlikley to be accessible to the LHC. [Since these figs are using 100/fb, with higher luminosity the reach may go somewhat beyond
250 GeV but for $kL \gtrsim 10$, much heavier masses will be quite difficult].

Given the many important questions in the SM that still remain unanswered, the currently observed paucity of new physics signals,  most likely,  just means that the relevant scales are  higher than the reach of the LHC runs at 7 or 8 TeV.
This may of course change in the next few years as LHC starts to run at higher energy $\approx$ 14 TeV. 
At least for the case of the very attractive RS models, 14 TeV is unlikely to be enough as suggested in
the preceding discussion.

\subsection {Lessons from our past: I. $\nu$ Oscillations}

At first sight it seems the properties of the Higgs-like scalar (~125 GeV) are quite consistent with the SM expectations.
In the context of RS models it means that the KK-scale is very likely heavier than ~3 TeV so it will be difficult to get
experimental verification at the LHC. To some degree, the RS models had been amended to try to
accomodate such a light scale with the desire (at least in part)  that the lower mass scale will be relevant at LHC;
of course the lower mass scale also ameliorates the amounting of tuning that is needed. However, from
flavor constraints and/or simplicity of the RS models,   scale less than ~10 TeV is quite difficult. Verification of these interesting ideas then will take additional work and a more powerful machine, perhaps O(100 TeV) cm energy.

In this context it may be worthwhile to remember that we encountered something analogous
in the search for neutrino mass and oscillation. In the 70's and early 80's experimental constraints over and over showed
that $\Delta m_{\nu}^2 \lesssim1 eV^2$. But while there were often debates and doubts,  the community did not abandon these searches as it was felt that 0 mass
neutrino is difficult to understand as we knew of no symmetry forcing $m_{\nu} = 0$. Indeed 
it took another decade or more of hard work and reduction of the bounds to  $\Delta m_{\nu}^2 \lesssim 10^{-4} eV^2$ before
neutrino oscillations were finally discovered!.

While understanding of Electroweak-Planck hierarchy may well be a technical issue,  
 occurrence of the replicas of flavors is  a well established fact
and understanding these is a pressing issue. RS gives a nice geomoetric interpretation.
To hunt for its experimental evidence we may well need to go to higher energy collider 
with energy appreciably higher than the LHC (14 TeV).

\subsection {Lessons from our past: II. $\epsilon_K$}

The point of view being espoused here is that the underlying BSM framework should give some understanding 
 of flavors (``Who ordered the muon?" );
flavors are an established experimental fact and therefore represent
 a much more tangible and pressing  issue than ``naturalness" or hierarchy.
Efforts at hadron collider  always need to be complemented by intensity/sensitivity 
frontier. The current tests of the SM-CKM paradigm are only around 10-15\%. 
It is not very sensible not to pursue more stringent and accurate
tests. Recall the indirect CP violation parameter  for $K_L \to \pi \pi$, $\epsilon_K \approx 
10^{-3}$!. History of Particle Physics would have been completely different if experiments seeking
kaon CP had been stopped even at O( 1\%).

Concurrent with efforts at a very high energy collider, it is imperative that more precise 
tests of the SM in the realm of ``null" tests~\cite{TGAS06} for CP violation, for neutron edm , for lepton
flavor violation etc.  be pursued with as high a priority as possible.

\subsection{Acknowledgements}
I must thank  my collaborators Hooman Davoudiasl and Thomas McElmurry for papers that form the backbone
of the work being reported here. I also want to thank Aleksandr Azatov and  Hying-Jin Kim. This research was supported in part by US
DOE grant No. DE-AC02-98CH10886.

\end{document}